\documentclass[12pt,twocolumn,A4]{article}

\usepackage{graphics}
\begin{document}
\onecolumn
\baselineskip 0.25in
\title{\Huge{Quantum Transport in Bridge Systems}}
\author{\Large {Santanu K. Maiti} \\ \\
        \Large {E-mail: {\em santanu.maiti@saha.ac.in}} \\ \\
        \Large {$^1$Theoretical Condensed Matter Physics Division} \\
        \Large {Saha Institute of Nuclear Physics} \\
        \Large {1/AF, Bidhannagar, Kolkata-700 064, India} \\ \\
        \Large {$^2$Department of Physics} \\
        \Large {Narasinha Dutt College} \\
        \Large {129, Belilious Road, Howrah-711 101, India}}
\date{}
\maketitle
\newpage
\tableofcontents

\newpage
\begin{center}
\addcontentsline{toc}{section}{\bf {Abstract}}
{\Large \bf Abstract}
\end{center}

\noindent
We study electron transport properties of some molecular wires and a 
unconventional disordered thin film within the tight-binding framework 
using Green's function technique. We show that electron transport is
significantly affected by quantum interference of electronic wave functions,
molecule-to-electrode coupling strengths, length of the molecular wire and
disorder strength. Our model calculations provide a physical insight to
the behavior of electron conduction across a bridge system.

\vskip 1in
\begin{flushleft}
\end{flushleft}

\newpage
\section{Introduction}

Recent advances in nanoscience and technology have made feasible to 
growth nanometer sized systems like, quantum wires~\cite{orella1,orella2,
til}, quantum dots~\cite{cron,hol1,hol2,shan}, molecular wires~\cite{yan}, 
etc. Quantum transport in such systems provides several novel features 
due to their reduced dimensionality and lateral quantum confinement.
The geometrical sensitivity of low-dimensional systems makes them truly
unique in offering the possibility of studying quantum transport in a 
very tunable environment. In the present age, designing of electronic 
circuits using a single molecule or a cluster of molecules becomes much
more widespread since the molecules are the fundamental building blocks
for future generation of electronic devices where electron transmits 
coherently~\cite{nitzan1,nitzan2}. Based on the pioneering work of 
Aviram and Ratner~\cite{aviram} where an innovative idea of a molecular 
electronic device was predicted for the first time, the development 
of a theoretical description of molecular devices has been pursued.
Later, many experiments~\cite{tali,metz,fish,reed1,reed2} have been 
carried out in different molecular bridge systems to justify the basic
mechanisms underlying such transport. Though there exists a vast literature
of theoretical as well as experimental study on electron transport in 
bridge systems, but yet the complete knowledge of conduction mechanism 
in such systems is not very well established even today. Many significant 
factors are there which can control the electron transport across a bridge 
system, and all these effects have to be taken into account properly to 
characterize such transport. For our illustrative purposes, here we 
mention very briefly some of them as follows. (I) The molecular coupling 
with side attached electrodes and the electron-electron 
correlation~\cite{tom} provide important signatures in the electron
transport. The understanding of the molecular coupling to the electrodes 
under non-equilibrium condition is a major challenge in this particular 
study. (II) The molecular geometry itself has a typical role. To emphasize 
it, Ernzerhof {\em et al.}~\cite{ern2} have predicted several model 
calculations and provided some new interesting results. (III) The quantum 
interference effect~\cite{mag,lau,baer1,baer2,baer3,tagami,walc1,gold,ern1} 
of electron waves passing through a bridge system probably the most important
aspect for controlling the electron transport, and a clear idea about it
is needed to reveal the transport mechanism. (IV) The dynamical fluctuation 
in the small-scale devices is another important factor which plays an
active role and can be manifested through the measurement of {\em shot
noise}, a direct consequence of the quantization of charge. It can be
used to obtain information on a system which is not available directly
through the conductance measurements, and is generally more sensitive
to the effects of electron-electron correlations than the average
conductance~\cite{blanter,walc2}. Beside these, several other
factors are there which may control the electron transport in a 
bridge system.

There exist several {\em ab initio} methods for the calculation of
conductance~\cite{yal,ven,xue,tay,der,dam,cheng1,cheng2,ern3,zhu1,zhu2} 
through a molecular bridge system. At the same time, tight-binding 
model has been extensively studied in the literature, and it has also 
been extended to DFT transport calculations~\cite{elst,fra}. The study 
of static density functional theory (DFT)~\cite{kohn1,kohn2} within 
the local-density approximation (LDA) to 
investigate the electron transport through nanoscale conductors, like 
atomic-scale point contacts, has met with great success. But when this 
similar theory applies to molecular junctions, theoretical conductances 
achieve much larger values compared to the experimental predictions, and 
these quantitative discrepancies need extensive and proper study in this 
particular field. In a recent work, Sai {\em et al.}~\cite{sai} have
predicted a correction to the conductance using the time-dependent
current-density functional theory since the dynamical effects give 
significant contribution in the electron transport, and illustrated some 
important results with specific examples. Similar dynamical effects have 
also been reported in some other recent papers~\cite{bush,ven1}, where 
authors have abandoned the infinite reservoirs, as originally introduced 
by Landauer, and considered two large but finite oppositely charged 
electrodes connected by a nanojunction. In this dissertation, we reproduce 
an analytic approach based on the tight-binding model to characterize the 
electron transport properties through some bridge systems, and utilize a 
simple parametric approach~\cite{muj1,muj2,sam,hjo,san1,san2,san3,san4} 
for these calculations. The model calculations are motivated by the fact 
that the {\em ab initio} theories are computationally much more expensive, 
while the model calculations by using the tight-binding formulation are 
computationally very cheap, and also provide a physical insight to the 
behavior of electron conduction through such bridge systems.

This dissertation can be organized in this way. Following the introductory
part (Section $1$), in Section $2$ we illustrate very briefly the 
methodology for the calculation of transmission probability, conductance 
and current through a finite size conductor attached to two metallic 
electrodes by using Green's function formalism. Section $3$ describes 
electron transport in some molecular wires. In Section $4$, we focus 
our study on electron transport through a unconventional disordered thin 
film in which disorder strength varies smoothly from layer to layer with 
the distance from its surface. Finally, we conclude our results 
in Section $5$.

\section{Theoretical Description}

\begin{figure}[ht]
{\centering \resizebox*{9cm}{1.7cm}{\includegraphics{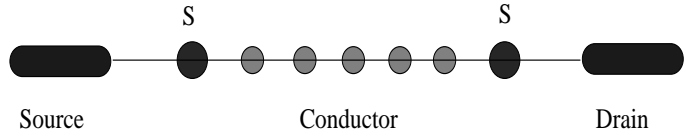}}
\par}
\caption{Schematic view of a finite size conductor attached to two metallic
electrodes, viz, source and drain through the lattice sites $S$ and $S$.}
\label{bridge}
\end{figure}

This section follows the methodology for the calculation of the
transmission probability ($T$), conductance ($g$) and current ($I$)
through a finite size conductor attached to two one-dimensional 
semi-infinite metallic electrodes by using Green's function technique. 
Let us refer to Fig.~\ref{bridge}, where a finite size conductor is 
attached to two metallic electrodes, viz, source and drain through 
the lattice sites $S$ and $S$.

At sufficient low temperature and bias voltage, we use the Landauer
conductance formula~\cite{datta,marc} to calculate the conductance $g$
of the conductor which can be expressed as,
\begin{equation}
g=\frac{2e^2}{h} T
\label{equ26}
\end{equation}
where $T$ becomes the transmission probability of an electron through 
the conductor. It can be expressed  
in terms of the Green's function of the conductor and its coupling to the
two electrodes by the relation~\cite{datta,marc},
\begin{equation}
T=Tr\left[\Gamma_S G_{c}^r \Gamma_D G_{c}^a\right]
\label{equ27}
\end{equation}
where $G_{c}^r$ and $G_{c}^a$ are respectively the retarded and advanced
Green's functions of the conductor including the effects of the electrodes.
The parameters $\Gamma_S$ and $\Gamma_D$ describe the coupling of the 
conductor to the source and drain respectively, and they can be defined 
in terms of their self-energies. For the complete system i.e., the conductor 
with the two electrodes the Green's function is defined as,
\begin{equation}
G=\left(\epsilon-H\right)^{-1}
\label{equ28}
\end{equation}
where $\epsilon=E+i\eta$. $E$ is the injecting energy of the source electron
and $\eta$ gives an infinitesimal imaginary part to $\epsilon$. Evaluation
of this Green's function requires the inversion of an infinite matrix as the
system consists of the finite conductor and the two semi-infinite electrodes.
However, the entire system can be partitioned into sub-matrices corresponding
to the individual sub-systems and the Green's function for the conductor
can be effectively written as,
\begin{equation}
G_c=\left(\epsilon-H_c-\Sigma_S-\Sigma_D\right)^{-1}
\label{equ29}
\end{equation}
where $H_c$ is the Hamiltonian of the conductor which can be written in the
tight-binding model within the non-interacting picture like,
\begin{equation}
H_c=\sum_i \epsilon_i c_i^{\dagger} c_i + \sum_{<ij>}t
\left(c_i^{\dagger}c_j + c_j^{\dagger}c_i \right)
\label{equ30}
\end{equation}
where $\epsilon_i$'s are the site energies and $t$ is the hopping strength
between two nearest-neighbor atomic sites in the conductor. Similar kind
of tight-binding Hamiltonian is also used to describe the two semi-infinite
one-dimensional perfect electrodes where the Hamiltonian is parametrized 
by constant on-site potential $\epsilon_0$ and nearest neighbor hopping 
integral $t_0$. In Eq.~(\ref{equ29}), $\Sigma_S=h_{Sc}^{\dagger}g_S h_{Sc}$ 
and $\Sigma_D=h_{Dc} g_D h_{Dc}^{\dagger}$ are the self-energy operators 
due to the two electrodes, where $g_S$ and $g_D$ correspond to the Green's
functions of the source and drain respectively. $h_{SC}$ and $h_{DC}$ are 
the coupling matrices and they will be non-zero only for the adjacent 
points of the conductor, $S$ and $S$ as shown in Fig.~\ref{bridge}, and 
the electrodes, respectively. The matrices $\Gamma_S$ and $\Gamma_D$ can 
be calculated through the expression,
\begin{equation}
\Gamma_{S(D)}=i\left[\Sigma_{S(D)}^r-\Sigma_{S(D)}^a\right]
\label{equ31}
\end{equation}
where $\Sigma_{S(D)}^r$ and $\Sigma_{S(D)}^a$ are the retarded and 
advanced self-energies respectively, and they are conjugate with each other.
Datta {\em et. al.}~\cite{tian} have shown that the self-energies can be
expressed like as,
\begin{equation}
\Sigma_{S(D)}^r=\Lambda_{S(D)}-i \Delta_{S(D)}
\label{equ32}
\end{equation}
where $\Lambda_{S(D)}$ are the real parts of the self-energies which
correspond to the shift of the energy eigenvalues of the conductor and 
the imaginary parts $\Delta_{S(D)}$ of the self-energies represent the
broadening of these energy levels. This broadening is much larger than the
thermal broadening and this is why we restrict our all calculations only
at absolute zero temperature. All the informations about the
conductor-to-electrode coupling are included into these two self-energies
as stated above and are described through the use of Newns-Anderson
chemisorption theory~\cite{muj1,muj2}. The detailed description of
this theory is available in these two references.
By utilizing the Newns-Anderson type model, we can express the conductance
in terms of the effective conductor properties multiplied by the effective
state densities involving the coupling. This allows us to study directly the
conductance as a function of the properties of the electronic structure of
the conductor within the electrodes.

The current passing across the conductor is depicted as a single-electron
scattering process between the two reservoirs of charge carriers. The
current $I$ can be computed as a function of the applied bias voltage $V$
through the relation~\cite{datta},
\begin{equation}
I(V)=\frac{e}{\pi \hbar}\int_{E_F-eV/2}^{E_F+eV/2} T(E,V) dE
\label{equ34}
\end{equation}
where $E_F$ is the equilibrium Fermi energy. For the sake of simplicity,
we assume that the entire voltage is dropped across the conductor-electrode
interfaces and this assumption doesn't greatly affect the qualitative aspects
of the $I$-$V$ characteristics. Such an assumption is based on the fact that,
the electric field inside the conductor especially for short conductors seems
to have a minimal effect on the conductance-voltage characteristics. On the
other hand, for quite larger conductors and high bias voltages the electric
field inside the conductor may play a more significant role depending on the
internal structure and size of the conductor~\cite{tian}, yet the effect
is quite small.

\section{Quantum Transport in Molecular Wires}

In this section, we narrate electron transport properties of some 
molecular wires consisting with polycyclic hydrocarbon molecules. 
These molecules are named as benzene, napthalene, anthracene and 
tetracene respectively. The transport properties in the molecular 
wires are significantly affected by the (i) quantum interference 
effects, (ii) molecule-to-electrode coupling strength, and (iii) 
length of the molecular wire, and here we discuss our results
in these aspects.

\subsection{Model}

\begin{figure}[ht]
{\centering \resizebox*{8.5cm}{8.5cm}{\includegraphics{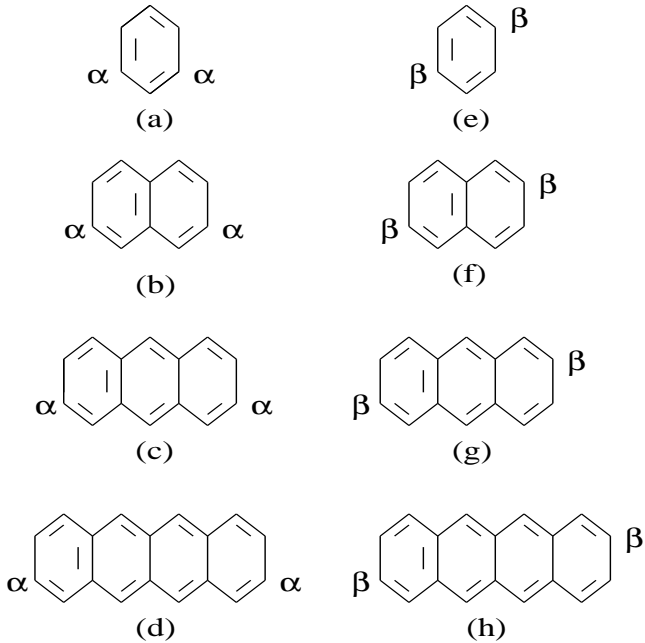}}\par}
\caption{Molecular model for the four different polycyclic hydrocarbon 
molecules. The molecules are benzene (one ring), napthalene (two rings), 
anthracene (three rings) and tetracene (four rings) respectively. These 
molecules are attached to the electrodes, at the $\alpha$-$\alpha$ 
positions called the cis configuration, and at the $\beta$-$\beta$ 
positions called the trans configuration via thiol (SH) groups.}
\label{hydro}
\end{figure}

In Fig.~\ref{hydro}, we show the model of the four different 
polycyclic hydrocarbon molecules. To reveal the quantum interference
effects, we consider two different arrangements of the molecular wires.
In one case, the molecules are attached to the electrodes at the
$\alpha$-$\alpha$ sites (see the first column of Fig.~\ref{hydro}). 
This is so-called the cis configuration. In the other case,
the electrodes are attached to these molecules at the $\beta$-$\beta$
sites, as presented in the second column of Fig.~\ref{hydro}. This
particular arrangement is so-called the trans configuration. In actual 
experimental set-up, the electrodes made from gold (Au) are used and 
the molecule coupled to the electrodes through thiol (SH) groups in 
the chemisorption technique where hydrogen (H) atoms remove and sulfur 
(S) atoms reside. To describe the polycyclic hydrocarbon molecules here
we use the similar kind of non-interacting tight-binding Hamiltonian 
as illustrated in Eq.~(\ref{equ30}).

\subsection{Results and Discussion}

Here we describe all the essential features of the electron transport
for the two distinct regimes. One is  so-called the weak coupling regime,
defined by the condition $\tau_{\{S,D\}} << t$. The other one is so-called
the strong-coupling regime, denoted by the condition $\tau_{\{S,D\}}\sim t$,
where $\tau_S$ and $\tau_D$ correspond to the hopping strengths of the 
molecule to the source and drain respectively. For these two limiting
cases we take the values of the different parameters as follows:
$\tau_S=\tau_D=0.5$, $t=2.5$ (weak-coupling) and $\tau_S=\tau_D=2$, 
$t=2.5$ (strong-coupling). Here we set the on-site energy $\epsilon_0=0$ 
(we can take any constant value of it instead of zero, since it gives only 
the reference energy level) for the electrodes, and the hopping strength 
$t_0=4$ in the two semi-infinite metallic electrodes. For the sake of 
simplicity, we set the Fermi energy $E_F=0$.

\begin{figure}[ht]
{\centering \resizebox*{10cm}{10cm}{\includegraphics{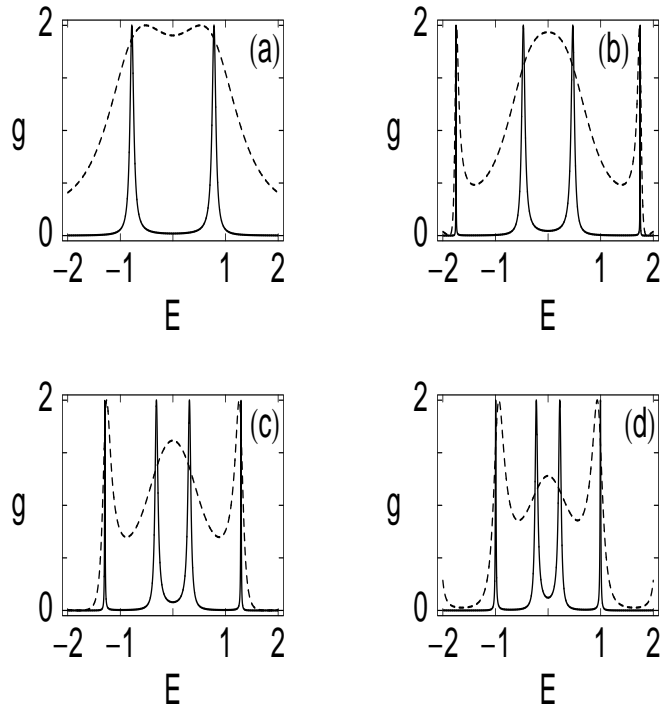}}\par}
\caption{$g$-$E$ characteristics of the molecular wires in the trans
configuration, where (a), (b), (c) and (d) correspond to the wires with 
benzene, napthalene, anthracene and tetracene molecules respectively. 
The solid and dotted curves represent the results in the weak and strong 
molecule-to-electrode coupling limits respectively.}
\label{transcond}
\end{figure}

Let us begin our discussion with the variation of the conductance $g$ as
a function of the injecting electron energy $E$. As representative
examples, in Fig.~\ref{transcond}, we plot the $g$-$E$ characteristics
for the molecular wires in which the molecules are attached to the 
electrodes in the trans configuration. Figures~\ref{transcond}(a), (b), 
(c) and (d) correspond to the results for the wires with benzene, 
napthalene, anthracene and tetracene molecules respectively. The solid 
and dotted curves represent the results in the weak and strong molecular 
coupling limits respectively. It is observed that, in the limit of
weak molecular coupling, the conductance shows very sharp resonance 
peaks for some
particular energy values, while almost for all other energies it ($g$) 
drops to zero. At these resonances, the conductance approaches the value 
$2$, and therefore, the transmission probability $T$ goes to unity since
we have the relation $g=2T$ from the Landauer conductance formula 
(see Eq.(\ref{equ26}) with $e=h=1$ in the present description). These
resonance peaks are associated with the energy eigenvalues of the single 
hydrocarbon molecules, and therefore we can say that the conductance 
spectrum manifests itself the electronic structure of the molecules. 
Now in the strong molecule-to-electrode coupling limit, all the resonances 
get substantial widths, which emphasize that the electron conduction takes 
place almost for all energy values. Such an enhancement of the resonance 
widths is due to the broadening of the molecular energy levels in the
limit of strong molecular coupling, where the contribution comes from 
the imaginary parts of the self-energies $\Sigma_S$ and 
$\Sigma_D$~\cite{datta} as mentioned earlier in the previous section.

\begin{figure}[ht]
{\centering \resizebox*{10cm}{10cm}{\includegraphics{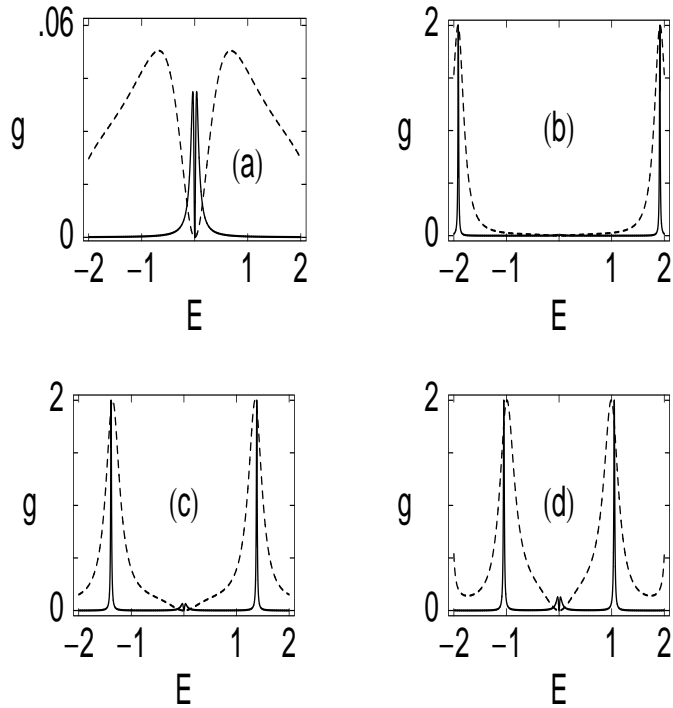}}\par}
\caption{$g$-$E$ characteristics of the molecular wires in the cis 
configuration, where (a), (b), (c) and (d) correspond to the wires with
benzene, napthalene, anthracene and tetracene molecules respectively.
The solid and dotted curves represent the results in the weak and 
strong molecule-to-electrodes coupling limits respectively.}
\label{ciscond}
\end{figure}

To illustrate the quantum interference effects on electron transport, 
in Fig.~\ref{ciscond}, we plot the conductance-energy ($g$-$E$) 
characteristics for the molecular wires where the molecules are attached 
to the electrodes
in the cis configuration. Figures~\ref{ciscond}(a), (b), (c) and (d)
correspond to the results of the wires with benzene, napthalene,
anthracene and tetracene molecules respectively. The solid and dotted 
lines indicate the same meaning as in Fig.~\ref{transcond}. These results 
predict that, some of the conductance peaks do not reach to unity anymore, 
and get much reduced value. This behavior can be understood in this way.
During the motion of the electrons from the source to the drain through
the molecules, the electron waves propagating along the different
possible pathways can get a phase shift among themselves according to
the result of quantum interference.
Therefore, the probability amplitude of getting the electron across the 
molecules either becomes strengthened or weakened. This causes the 
transmittance cancellations and provides anti-resonances in the 
conductance spectrum. Thus it can be emphasized that the electron 
transmission is strongly affected by the quantum interference effects 
and hence the molecule to electrodes interface structures.

\begin{figure}[ht]
{\centering \resizebox*{9cm}{10cm}{\includegraphics{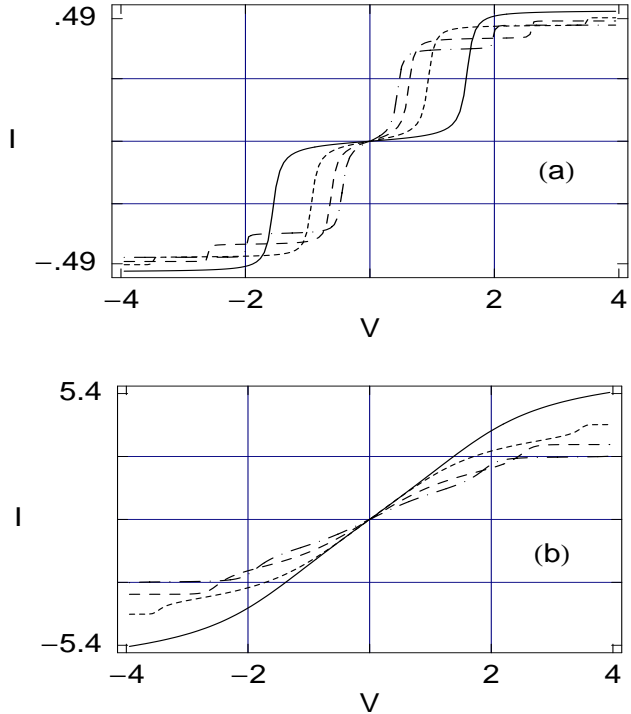}}\par}
\caption{$I$-$V$ characteristics of the molecular wires in the trans
configuration, where the solid, dotted, dashed and dot-dashed curves
correspond to the results for the wires with benzene, napthalene, 
anthracene and tetracene molecules respectively. (a) weak-coupling
limit and (b) strong-coupling limit.}
\label{transcurr}
\end{figure}

The scenario of the electron transfer through the molecular junction
becomes much more clearly visible by investigating the current-voltage 
($I$-$V$) characteristics. The current through the molecular systems 
can be computed by the integration procedure of the transmission 
function $T$ (see Eq.(\ref{equ34})), where the function $T$ varies 
exactly similar to the conductance spectra, differ only in magnitude 
by the factor $2$, since the relation $g=2T$ holds from the Landauer
conductance formula (Eq.(\ref{equ26})). To reveal this fact, in
Fig.~\ref{transcurr} we plot the current-voltage characteristics
for the molecular wires in which the molecules attached to the
electrodes in the trans configuration. Figures~\ref{transcurr}(a)
and (b) correspond to results for the weak- and strong-coupling
limits respectively. The solid, dotted, dashed and dot-dashed curves
represent the variations of the currents with the bias voltage $V$
for the molecular wires consisting with benzene, napthalene,
anthracene and tetracene molecules respectively.
In the weak molecular coupling, the current exhibits staircase-like 
structure with fine steps as a function of the applied bias voltage. 
This is due to the existence of the sharp resonance peaks in the 
conductance spectra in this limit of coupling, since the current is 
computed by the integration method of the transmission function $T$. 
With the increase of the applied bias voltage, the electrochemical 
potentials on the electrodes are shifted gradually, and finally cross 
one of the quantized energy levels of the molecule. Therefore,
a current channel is opened up and the current-voltage characteristic
curve provides a jump. The other important feature is that the threshold
bias voltage of the electron conduction across the wire significantly
depends on the length of the wire in this weak-coupling limit. 
On the other hand, for the strong molecular coupling, the current 
varies almost continuously with the applied bias voltage and achieves 
much large amplitude than the weak-coupling case. This is because the 
resonance peaks get broadened due to the broadening of the energy 
levels in the strong-coupling limit which provide much larger current 
amplitude as we integrate the transmission function $T$ to get the 
current. Thus by tuning the molecule-to-electrode coupling, one can 
achieve very high current from the very low one. For this strong-coupling
limit, the electron starts to conduct as long as the bias voltage is
applied, in contrary to that of the weak-coupling case, for all these
molecular wires. Thus we can say that, for this strong molecular 
coupling limit, the threshold bias voltage of the electron conduction 
is almost independent of the length of the molecular wire.

The effects of the quantum interference on electron transport can be
much more clearly understood from the current-voltage characteristics
\begin{figure}[ht]
{\centering \resizebox*{9cm}{10cm}{\includegraphics{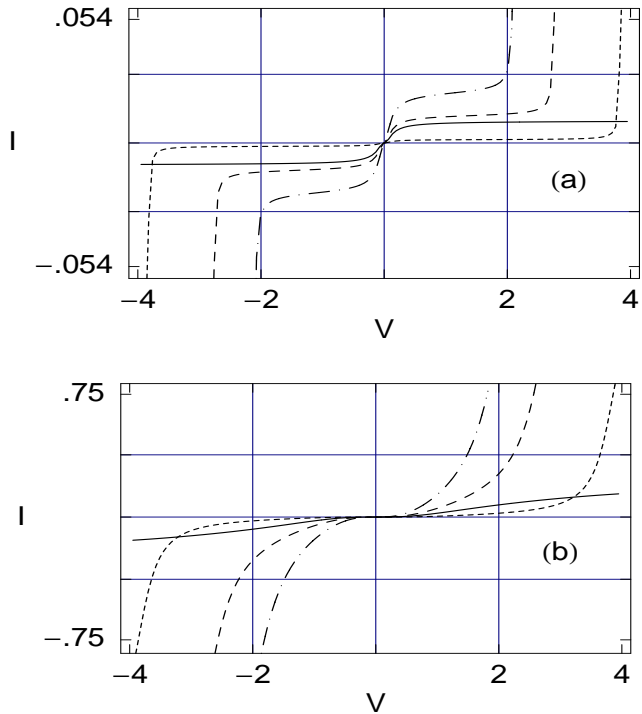}}\par}
\caption{$I$-$V$ characteristics of the molecular wires in the cis
configuration, where the solid, dotted, dashed and dot-dashed curves
correspond to the results for the wires with benzene, napthalene, 
anthracene and tetracene molecules respectively. (a) weak-coupling
limit and (b) strong-coupling limit.}
\label{ciscurr}
\end{figure}
plotted in Fig.~\ref{ciscurr}. In this case, the molecular wires are
attached to the electrodes in the cis configuration, where 
Figs.~\ref{ciscurr}(a) and (b) correspond to the results for the weak-
and strong-coupling limits respectively. The solid, dotted, dashed and
dot-dashed curves represent the same meaning as in Fig.~\ref{transcurr}.
Our results show that, for these wires the current amplitudes get reduced 
enormously compared to the results obtained for the wires when the
molecules are attached with the electrodes in the trans configuration.
This is solely due to the quantum interference effects among all the 
possible pathways that the electron can take. Therefore, we can predict 
that designing a molecular device is significantly influenced by the 
quantum interference effects i.e., the molecule to electrodes interface 
structures.

In conclusion of this section, we have introduced a parametric approach 
based on the tight-binding model to investigate the electron transport 
properties in some polycyclic hydrocarbon molecules attached to two 
semi-infinite one-dimensional metallic electrodes. This technique may 
be utilized to study the electronic transport in any complicated 
molecular bridge system. The conduction of electron through the 
hydrocarbon molecules is strongly influenced by the molecule-to-electrode 
coupling strength, length of the molecule, and the quantum interference 
effects. This study reveals that designing a whole system that includes
not only the molecule but also the molecule-to-electrode coupling and the
interface structures are highly important in fabricating molecular electronic
devices.

\section{Quantum Transport in a Thin Film}

Here we explore a novel feature of electron transport in a unconventional 
disordered thin film where disorder strength varies smoothly from its 
surface. In the present age of nanoscience and technology, it becomes 
quite easy to fabricate a nano-scale device where charge carriers are 
scattered mainly from its surface boundaries~\cite{kou,zho1,zho2,ding1,
ding2,san5,san6,san7,san8,san9}, and not from the inner core region. 
It is completely opposite to that of a traditional doped system where 
the dopant atoms are 
distributed uniformly along the system. For example, in shell-doped 
nanowires the dopant atoms are spatially confined within a few atomic 
layers in the shell region of a nanowire. In such a shell-doped nanowire, 
Zhong and Stocks~\cite{zho1} have shown that the electron dynamics 
undergoes a localization to quasi-delocalization transition beyond some 
critical doping. In other very recent work~\cite{ding1}, Yang {\em et al.} 
have also observed such a transition in edge disordered graphene 
nanoribbons upon varying the strength of edge disorder. From extensive 
studies of electron transport in such unconventional systems, it has been 
suggested that the surface states~\cite{yu}, surface scattering~\cite{cui} 
and the surface reconstructions~\cite{rurali} may be responsible to exhibit 
several diverse transport properties. Motivated with these systems, here 
we focus our study of electron transport in a special type of thin film, 
in which disorder strength varies smoothly from layer to layer with the 
distance from its surface. This system shows a peculiar behavior of 
electron transport where the current amplitude increases with the increase 
of the disorder strength in the limit of strong disorder, while it decreases 
in the weak disorder limit. On the other hand, for the traditional 
disordered thin film i.e., the film subjected to uniform disorder, the 
current amplitude always decreases with the increase of the disorder 
strength.

\subsection{Model}

\begin{figure}[ht]
{\centering \resizebox*{9cm}{4cm}{\includegraphics{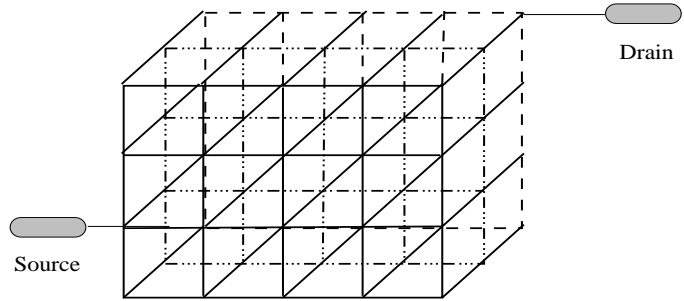}}\par}
\caption{Schematic view of a smoothly varying disordered thin film attached
to two metallic electrodes (source and drain). The top most front layer
(solid line) is the highest disordered layer and the disorder strength
decreases smoothly to-wards the bottom layer keeping the lowest bottom
layer (dashed line) as disorder free. Two electrodes are attached at the
two extreme corners of the bottom layer.}
\label{quantumfilm}
\end{figure}

Let us refer to Fig.~\ref{quantumfilm}, where a thin film is attached to 
two metallic electrodes, viz, source and drain. In this film, disorder 
strength varies smoothly from the top most disordered layer (solid line) 
to-wards the bottom layer, keeping the lowest bottom layer (dashed line) 
as disorder free. The electrodes are symmetrically attached at the two 
extreme corners of the bottom layer. 
Both this film and the two side attached electrodes are described by the
similar kind of tight-binding Hamiltonian as prescribed in Eq.~(\ref{equ30}).
Now to achieve our required unconventional thin film, we choose the site 
energies ($\epsilon_i$'s in Eq.~(\ref{equ30})) randomly from a ``Box" 
distribution function such that the top most front layer becomes the 
highest disordered layer with strength $W$, and the strength of disorder 
decreases smoothly to-wards the bottom layer as a function of $W/(N_l-m)$, 
where $N_l$ gives the total number of layers and $m$ represents the total 
number of ordered layers from the bottom side of the film. On the other 
hand, in the conventional disordered thin film, all the layers are
subjected to the same disorder strength $W$.

Here, we concentrate our study on the determination of the typical current 
amplitude which is obtained from the relation,
\begin{equation}
I_{typ}=\sqrt{<I^2>_{W,V}}
\label{equ6}
\end{equation}
where $W$ and $V$ correspond to the impurity strength and the applied
bias voltage respectively.

\subsection{Results and Discussion}

All the numerical calculations we present here are performed for some
particular values of the different parameters, and all the basic
features remain also invariant for some the other parametric values. 
The values of the required parameters are as follows. The coupling 
strengths of the film to the electrodes are taken as $\tau_S=\tau_D=1.5$, 
the nearest-neighbor hopping integral in the film is fixed to $t=1$.
The on-site potential and the hopping integral in the electrodes are
set as $\epsilon_0=0$ and $t_0=2$ respectively. In addition to these,
here we also introduce another three parameters $N_x$, $N_y$ and $N_z$ to
specify the system size of the thin film, where they correspond to the
total number of lattice sites along the $x$, $y$ and $z$ directions
of the film respectively. In our numerical calculations, the typical
current amplitude ($I_{typ}$) is determined by taking the average
over the disordered configurations and bias voltages (see Eq.(\ref{equ6})).
Since in this particular model the site energies are chosen randomly,
we compute $I_{typ}$ by taking the average over a large number ($60$)
of disordered configurations in each case to get much accurate result.
On the other hand, for the averaging over the bias voltage $V$, we set
the range of it from $-10$ to $10$. In this presentation, we focus
only on the systems with small sizes since all the qualitative behaviors
remain also invariant even for the large systems.

\begin{figure}[ht]
{\centering \resizebox*{8cm}{5cm}{\includegraphics{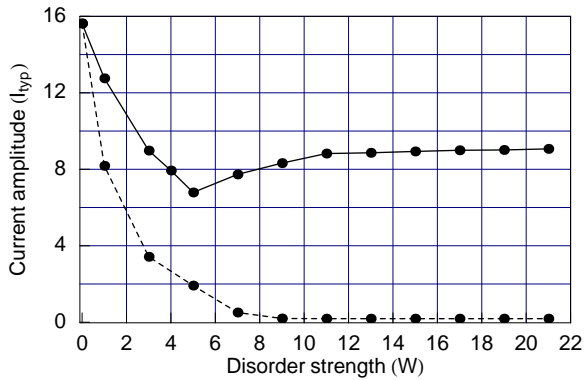}}\par}
\caption{$I_{typ}$ vs. $W$ for the two different types of thin films 
with $N_x=10$, $N_y=8$ and $N_z=5$. Here we set $m=1$. The solid and dotted 
curves correspond to the smoothly varying and complete disordered films 
respectively.}
\label{disorder2}
\end{figure}

Figure~\ref{disorder2} represents the variation of the typical current
amplitude ($I_{typ}$) as a function of disorder ($W$) for some typical
thin films with $N_x=10$, $N_y=8$ and $N_z=5$. Here we set $m=1$, i.e., 
only the lowest bottom layer of the unconventional disordered thin film 
is free from any disorder.
The solid and dotted curves correspond to the results of the smoothly
varying and complete disordered
thin films respectively. A remarkably different behavior is observed
for the smoothly varying disordered film compared to the film with
complete disorder. In the later system, it is observed that $I_{typ}$
decreases rapidly with $W$ and eventually it drops to zero for the
higher value of $W$. This reduction of the current is due to the fact 
that the eigenstates become more localized~\cite{lee} with the increase
of disorder, and it is well established from the theory of Anderson
localization~\cite{anderson}. The appreciable change in the variation 
of the typical current amplitude takes place only for the unconventional
disordered film. In this case, the current amplitude decreases
initially with $W$ and after reaching to a minimum at $W=W_c$ (say),
it again increases. Thus the anomalous behavior is observed beyond the
critical disorder strength $W_c$, and we are interested particularly
in this regime where $W>W_c$. In order to illustrate this peculiar
behavior, we consider the smoothly varying disordered film as a coupled
system combining two sub-systems. The coupling exists between the
lowest bottom ordered layer and the other disordered layers. Thus
the system can be treated, in other way, as a coupled order-disorder
separated thin film. For this coupled system we can write the
Schr\"{o}dinger equations as: $(H_0-H_1)\psi_0=E\psi_0$ and
$(H_d-H_2)\psi_d=E\psi_d$. Here $H_0$ and $H_d$ represent the
sub-Hamiltonians of the ordered and disordered regions of the film
respectively, and $\psi_0$ and $\psi_d$ are the corresponding
eigenfunctions. The terms $H_1$ and $H_2$ in the above two
expressions are the most significant and they can be expressed as:
$H_1=H_{od}(H_d-E)^{-1}H_{do}$ and $H_2=H_{do}(H_o-E)^{-1}H_{od}$.
$H_{od}$ and $H_{do}$ correspond to the coupling between the ordered
region and the disordered region~\cite{zho1,zho2}. From these 
mathematical expressions, the anomalous behavior of the electron 
transport in the film
\begin{figure}[ht]
{\centering \resizebox*{8cm}{5cm}{\includegraphics{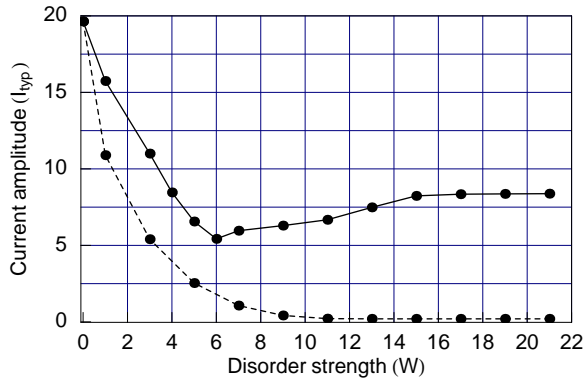}}\par}
\caption{$I_{typ}$ vs. $W$ for the two different types of thin films 
with $N_x=12$, $N_y=10$ and $N_z=6$. Here we set $m=2$. The solid and 
dotted curves correspond to the identical meaning as in 
Fig.~\ref{disorder2}.}
\label{disorder3}
\end{figure}
can be described clearly. In the absence of any interaction between
the ordered and disordered regions, we can assume the full system as
a simple combination of two independent sub-systems. Therefore, we get
all the extended states in the ordered region, while the localized
states are obtained in the disordered region. In this situation, the
motion of an electron in any one region is not affected by the other.
But for the coupled system, the motion of the electron is no more
independent, and we have to take the combined effects coming from both
the two regions. With the increase of disorder, the scattering
effect becomes dominated more, and thus the reduction of the current
is expected. This scattering is due to the existence of the localized
eigenstates in the disordered regions. Therefore, in the case of
strong coupling between the two sub-systems, the motion of the electron
in the ordered region is significantly influenced by the disordered
regions. Now the degree of this coupling between the two sub-systems
solely depends on the two parameters $H_1$ and $H_2$, those are expressed
earlier. In the limit of weak disorder, the scattering effect from both
the two regions is quite significant since then the terms $H_1$ and $H_2$
have reasonably high values. With the increase of disorder,
$H_1$ decreases gradually and for a very large value of $W$ it becomes
very small. Hence the term $(H_0-H_1)$ effectively goes to $H_0$ in the
limit $W \rightarrow 0$, which indicates that the ordered region becomes
decoupled from the disordered one. Therefore, in the higher disorder
regime the scattering effect becomes less significant from the ordered
region, and it decreases
with $W$. For the low regime of $W$, the eigenstates of both the two
effective Hamiltonians, $(H_0-H_1)$ and $(H_d-H_2)$, are localized.
With the increase of $W$, $H_1$ gradually decreases, resulting in
much weaker localization in the states of $(H_0-H_1)$, while the states
of $(H_d-H_2)$ become more localized. At a critical value of $W=W_c$
(say) ($\simeq$ band width of $H_0$), we get a separation between
the much weaker localized states and the strongly localized states.
Beyond this value, the weaker localized states become more extended
and the strongly localized states become more localized with the
increase of $W$. In this situation, the current is obtained mainly
from these nearly extended states which provide the larger current
with $W$ in the higher disorder regime.

To illustrate the size dependence of the film on the electron transport,
in Fig.~\ref{disorder3} we plot the variation of the typical current
amplitude for some typical thin films with $N_x=12$, $N_y=10$ and $N_z=6$. 
For these films we take $m=2$, i.e., two layers from the bottom side of
the smoothly varying disordered film are free from any disorder. The solid 
and dotted curves correspond to the identical meaning as in 
Fig.~\ref{disorder2}. For both the unconventional and traditional 
disordered films, we get almost the similar behavior of the current 
as described in Fig.~\ref{disorder2}. This study shows that the typical 
current amplitude strongly depends on the finite size of the thin film.

In summary of this section, we have provided a numerical study to 
exhibit the anomalous behavior of electron transport in a unconventional 
disordered thin film, where the disorder strength varies smoothly from 
its surface. Our numerical results have predicted that, in the smoothly 
varying disordered film, the typical current amplitude decreases with 
$W$ in the weak disorder regime ($W<W_c$), while it increases in the 
strong disorder regime ($W>W_c$). On the other hand for the conventional 
disordered film, the current amplitude always decreases with disorder. 
In this present investigations, we have also studied the finite size 
effects which reveal that the typical current amplitude strongly
depends on the size of the film. Similar type of anomalous quantum
transport can also be observed in lower dimensional systems like, edge
disordered graphene sheets of single-atom-thick, surface disordered finite
width rings, nanowires, etc. 

\section{Concluding Remarks}

In this dissertation, we have demonstrated the quantum transport properties
in different types of bridge systems like, molecular wires and thin films.
The physics of electron transport through these nanoscale systems is 
surprisingly rich. Many fundamental experimentally observed phenomena in
such systems can be understood by using simple arguments. In particular,
the formal relation between conductance and transmission coefficients (the
Landauer formula) has enhanced the understanding of electronic transport
in the bridge system. We have investigated the electron transport
properties of some molecular bridge systems and unconventional disordered
thin films within the tight-binding framework using Green's function 
technique and tried to explain how electron transport is affected by the
quantum interference of the electronic wave functions, molecule-to-electrode 
coupling strengths, length of the molecular wire and disorder strength. 
Our model calculations provide a physical insight to the behavior of 
electron conduction in these bridge systems.

First, we have studied the electron transport in some molecular wires
consisting with some polycyclic hydrocarbon molecules. Most interestingly,
it has been observed that the transport properties are significantly 
influenced by the molecular coupling strength to the side attached 
electrodes, quantum interference effects and length of the molecule.
Our study has emphasized that the molecule to electrodes interface 
structures are highly important in fabricating molecular electronic 
devices. Secondly, we have investigated the electron transport in a
unconventional disordered thin film. Most remarkably, we have noticed 
that the typical current amplitude increases with the disorder strength
in the strong disorder regime, while it decreases with the strength
of disorder in the weak disorder regime. This particular study has 
suggested that the carrier transport in an order-disorder separated 
mesoscopic device may be tailored to desired properties through doping 
for different applications.

\addcontentsline{toc}{section}{\bf {References}}

\end{document}